\documentclass{aastex}
\usepackage{emulateapj5}
\submitted{to appear in \textit{The Astronomical Journal}}
\lefthead{RED GALAXIES}
\righthead{HOGG ET AL}

\newcommand{\fixredshift}{0.1}
\newcommand{\fixmag}[1]{{^{\fixredshift}\!{#1}}}
\newcommand{\fixu}{\fixmag{u}}
\newcommand{\fixg}{\fixmag{g}}
\newcommand{\fixr}{\fixmag{r}}
\newcommand{\fixi}{\fixmag{i}}
\newcommand{\fixz}{\fixmag{z}}
\newcommand{\fixmur}{\mu_\fixr}

\newcommand{\fixMr}{M_\fixr}

\newcommand{\fixLr}{L_\fixr}
\newcommand{\fixLi}{L_\fixi}
\newcommand{\Lstar}{L^{\ast}}
\newcommand{\fixug}{\fixmag{(u\!-\!g)}}
\newcommand{\fixgr}{\fixmag{(g\!-\!r)}}

\newcommand{\Vmax}{V_\mmax}
\newcommand{\kcorrect}{\texttt{kcorrect} version \texttt{v1\_3}}
\newcommand{\lsssample}{\texttt{sample8}}

\newcommand{\mmin}{\mathrm{min}}
\newcommand{\mmax}{\mathrm{max}}
\newcommand{\minmax}{\mathrm{\left\{^{min}_{max}\right\}}}

\newcommand{\NYU}{1}
\newcommand{\hogg}{2}
\newcommand{\Princeton}{3}
\newcommand{\APO}{4}
\newcommand{\JHU}{5}
\newcommand{\Tokyo}{6}
\newcommand{\ICRR}{7}
\newcommand{\USNO}{8}
\newcommand{\Chicago}{9}
\newcommand{\Flagstaff}{10}
\newcommand{\CMU}{11}
\newcommand{\PSU}{12}

\begin{document}
\title{         The luminosity density of red galaxies}

\author{        David~W.~Hogg\altaffilmark{\NYU,\hogg},
                Michael~Blanton\altaffilmark{\NYU},
                Iskra~Strateva\altaffilmark{\Princeton},
                Neta~A.~Bahcall\altaffilmark{\Princeton},
                J.~Brinkmann\altaffilmark{\APO},
                Istvan~Csabai\altaffilmark{\JHU},
                Mamoru~Doi\altaffilmark{\Tokyo},
                Masataka~Fukugita\altaffilmark{\ICRR},
                Greg~Hennessy\altaffilmark{\USNO},
                \v{Z}eljko~Ivezi\'{c}\altaffilmark{\Princeton},
                G.~R.~Knapp\altaffilmark{\Princeton},
                Don~Q.~Lamb\altaffilmark{\Chicago},
                Robert~Lupton\altaffilmark{\Princeton},
                Jeffrey~A.~Munn\altaffilmark{\Flagstaff},
                Robert~Nichol\altaffilmark{\CMU},
                David~J.~Schlegel\altaffilmark{\Princeton},
                Donald~P.~Schneider\altaffilmark{\PSU},
                Donald~G.~York\altaffilmark{\Chicago}}

\altaffiltext{\NYU}{Center for Cosmology and Particle Physics,
                Department of Physics, New York University,
                4 Washington Place, New York, NY 10003}
\altaffiltext{\hogg}{\texttt{david.hogg@nyu.edu}}
\altaffiltext{\Princeton}{Department of Astrophysical Sciences,
                Princeton University, Princeton, NJ 08540}
\altaffiltext{\APO}{Apache Point Observatory,
                P.O. Box 59, Sunspot, NM 88349}
\altaffiltext{\JHU}{Department of Physics and Astronomy,
                The Johns Hopkins University,
                3701 San Martin Drive, Baltimore, MD 21218}
\altaffiltext{\Tokyo}{Institute of Astronomy, School of Science,
                University of Tokyo, Mitaka, Tokyo, 181-0015 Japan}
\altaffiltext{\ICRR}{Institute for Cosmic Ray Research,
                University of Tokyo,
                Kashiwa, Chiba, 277-8582 Japan}
\altaffiltext{\USNO}{United States Naval Observatory,
                3450 Massachusetts Avenue, NW, Washington, DC 20392}
\altaffiltext{\Chicago}{Enrico Fermi Institute, University of Chicago,
                5640 S. Ellis Ave., Chicago, IL 60637}
\altaffiltext{\Flagstaff}{United States Naval Observatory,
                Flagstaff Station,
                P.O. Box 1149, Flagstaff, AZ  86002}
\altaffiltext{\CMU}{Deptartment of Physics, Carnegie Mellon University,
                5000 Forbes Ave., Pittsburgh, PA 15232}
\altaffiltext{\PSU}{Department of Astronomy and Astrophysics,
                Pennsylvania State University, University Park, PA 16802}

\begin{abstract}
A complete sample of $7.7\times 10^4$ galaxies with five-band imaging
and spectroscopic redshifts from the Sloan Digital Sky Survey is used
to determine the fraction of the optical luminosity density of the
Local Universe (redshifts $0.02<z<0.22$) emitted by red galaxies.  The
distribution in the space of rest-frame color, central surface
brightness, and concentration is shown to be highly clustered and
bimodal; galaxies fall primarily into one of two distinct classes.
One class is red, concentrated and high in surface brightness; the
other is bluer, less concentrated, and lower in central surface
brightness.  Elliptical and bulge-dominated galaxies preferentially
belong to the red class.  Even with a very restrictive definition of
the red class that includes limits on color, surface brightness and
concentration, the class comprises roughly one fifth of the number
density of galaxies more luminous than $0.05\,\Lstar$ and produces two
fifths of the total cosmic galaxy luminosity density at
$0.7~\mathrm{\mu m}$.  The natural interpretation is that a large
fraction of the stellar mass density of the Local Universe is in very
old stellar populations.
\end{abstract}

\keywords{
  cosmology: observations
  ---
  galaxies: elliptical and lenticular, cD
  ---
  galaxies: evolution
  ---
  galaxies: fundamental parameters
  ---
  galaxies: stellar content
}

\section{Introduction}

The local Universe sparkles with bright blue stellar populations in
disk and irregular galaxies.  However, both age and dust (circumstellar or
interstellar) make a population dimmer and redder, so red populations
contain much more stellar mass per unit luminosity.  This makes it
remarkable that the most luminous (visible and near-infrared) galaxies
in the Universe are red galaxies (eg, Blanton et al 2001); their
dominance in luminosity represents enormous dominance in stellar mass.

The typical luminous red galaxies of the local Universe are
kinematically hot (ie, not rotationally supported), have smooth radial
profiles, have a narrow range of stellar mass-to-light ratios, show
high $\alpha$-element metallicities, are (relatively) iron-poor, and
reside preferentially in the Universe's most over-dense environments
(as reviewed by Kormendy \& Djorgovski 1989, and Roberts \& Haynes
1994).  They are red because their stellar populations are
old---comparable in age to the Universe itself---not because they are
obscured by dust.

It is now well established that the average cosmic star formation rate
per comoving volume has been declining from redshifts near unity to
the present day (eg, Pei \& Fall 1995, Lilly et al 1996, and many more
references cited in Hogg 2001).  This observed trend is strong enough
that the majority of the stars in the local Universe are expected to
be old ($>6$~Gyr).  This appears consistent with observations of the
distribution of stellar ages at the present day (eg, Fukugita et al
1998).  What is not currently agreed-upon is the detailed stellar age
distribution at old ages.  Direct measurements of the cosmic star
formation rate at very early times are hindered by selection effects
and the strong influence of dust; measurements of age distributions of
nearby stellar populations are confused by the asymptotic similarity
of spectral energy distributions with increasing age.

In this \textsl{Article} we look at the luminosity density in red
galaxies in the $z\sim 0.1$ local Universe using observations from the
Sloan Digital Sky Survey (SDSS).  We begin by suggesting a new method
for selecting red galaxies designed to avoid contamination by young
galaxies reddened by dust.  We find that the red galaxies contribute a
large fraction of the total galaxy luminosity density of the Universe
at $\lambda\sim 0.7~\mathrm{\mu m}$, such a large fraction that they
make up the majority of the cosmic stellar mass density.

In what follows, a cosmological world model with
$(\Omega_\mathrm{M},\Omega_\mathrm{\Lambda})=(0.3,0.7)$ is adopted.
The model has total age $9.4\,h^{-1}~\mathrm{Gyr}$, where $h$ is the
Hubble constant in units of $100~\mathrm{km\,s^{-1}\,Mpc^{-1}}$;
$h=0.67$ makes the total age $14~\mathrm{Gyr}$ (eg, Hogg 1999).

\section{The sample}

\subsection{Imaging and photometry}

The SDSS is taking $u$, $g$, $r$, $i$ and $z$-band drift-scan imaging
of the Northern Galactic Cap, and taking spectra of roughly $9\times
10^5$ galaxies in that region, most with $r<17.7~\mathrm{mag}$ in the
AB system (York et al 2000).  The photometric system and imaging
hardware are described in detail elsewhere (Fukugita et al 1996; Gunn
et al 1998; Smith, Tucker et al 2002).  An automated image-processing
system detects astronomical sources and measures many photometric
properties (Lupton et al 2001; Stoughton et al 2002, Pier et al 2002,
Lupton et al in preparation).  The photometric parameters of interest
for this study are as follows: the Petrosian (1976) radius
$\theta_\mathrm{pet}$ is the angular radius at which the mean surface
brightness of the source in the SDSS $r$-band image inside that radius
is five times higher than the mean surface brightness in a narrow
annulus centered on that radius.  The Petrosian fluxes (with
corresponding magnitudes $m_\mathrm{pet}$) are the total fluxes in the
five SDSS bandpass images within a circular aperture of radius
$2\,\theta_{r,\mathrm{pet}}$ (twice the Petrosian radius in the
$r$-band).  The half-light and 90-percent radii $\theta_{50}$ and
$\theta_{90}$ are the angular radii within which 50 and 90~percent of
the Petrosian flux is found.  All magnitudes are extiction-corrected
with IRAS-based dust maps (Schlegel Finkbeiner \& Davis 1998).

The median magnitude of a galaxy in our sample is $r=17~\mathrm{mag}$,
so it is detected in the imaging data at a signal-to-noise ratio
$S/N>300$; the photometric properties of the galaxies in the sample
are measured with high precision.  The accuracy is limited by
systematic errors at the few~percent to ten~percent level (Stoughton
et al 2002).

\subsection{Spectroscopy}

SDSS ``Main Sample'' galaxies are chosen as spectroscopic targets on
the basis of $r$-band magnitude, angular size (to remove stars),
half-light surface brightness (to ensure spectroscopic success), and
some positional constraints from the finite size of the fibers in the
fiber-fed multi-object spectrographs (Strauss et al 2002, Blanton et
al 2002a).

The SDSS spectra are at resolution 2000 with wavelength coverage
$3900<\lambda<9000$~\AA.  Typical spectra have signal-to-noise better
than 10 per resolution element at the central wavelengths.  For the
purpose of this study, the spectroscopic data are used only for
obtaining galaxy redshifts.  The SDSS spectroscopic reduction pipeline
(Schlegel et al in preparation) produces a one-dimensional spectrum
corresponding to each fiber. The SDSS has an official pipeline which
identifies redshifts (SubbaRao et al in preparation).  However, the
redshifts used in the current analysis are determined independently
using a separate pipeline (Schlegel et al in preparation) whose
results are nearly identical (at the 99\% level) for the main galaxy
sample to the official SDSS pipeline (which currently performs better
for objects with unusual spectra, such as certain types of stars and
QSOs).

\subsection{Correction to a fixed redshift frame}

Galaxies observed at different redshifts are observed through
bandpasses at different wavelengths in the galaxy's rest frame.  The
photometric measurements have been $K$-corrected to a fixed wavelength
frame using a $K$-correction pipeline (\kcorrect) described in detail
elsewhere (Blanton et al 2002b). The method is, briefly, as follows. A
linear combination of four SED templates is fit to the set of
broad-band magnitudes for each galaxy.  The four templates are chosen
to yield good fits to the broad-band magnitudes of the entire set of
galaxies in the SDSS spectroscopic sample. The fit provides a
correction from the rest-frame band at the redshift of the galaxy to
the rest-frame band at $z=0.1$.  Correction to the (roughly) median
redshift $z=0.1$ minimizes the typical correction magnitude.  The
resulting fixed-frame magnitude system is denoted ``$\fixu$, $\fixg$,
$\fixr$, $\fixi$, $\fixz$''.  The effective wavelengths are roughly
3220, 4340, 5660, 6940 and 8300~\AA.  The zeropoints are as close to
AB (Oke \& Gunn 1983) as is currently possible with SDSS calibration;
a hypothetical source with $f_\nu=3.63\times
10^{-23}~\mathrm{W\,m^{-2}\,Hz^{-1}}$ at all frequencies $\nu$ would
be close to $0~\mathrm{mag}$ in all five bands.

\subsection{Sample selection}

The sample used here is a subsample (known as ``\lsssample'' within
the SDSS collaboration) of spectroscopic targets selected from all
SDSS imaging targeted before 2001 June 20 and spectroscopically
observed before 2001 October 20.  A galaxy is included in our analysis
for this paper if it is a main sample galaxy (Strauss et al 2002b)
satisfying the following cuts: (1)~apparent magnitude $14.5 <
r_\mathrm{pet}<17.77~\mathrm{mag}$ (the exact faint limit is variable
at the 0.2~mag level, depending on the SDSS Main Sample photometric
selection in that region of sky, which has varied somewhat as the
survey has progressed); (2)~mean surface brightness
$\mu_r<24.5~\mathrm{mag}$ in $1~\mathrm{arcsec^2}$ within the
half-light radius (variable at the 0.5~mag level, depending again on
SDSS Main Sample selection); (3)~redshift $0.02<z<0.22$; and
(4)~fixed-frame absolute magnitude $-23.5<\fixMr<-17.5~\mathrm{mag}$,
corresponding to $0.05<(\fixLr/\Lstar)<12$ (Blanton et al 2001).

These cuts left a total sample of 77,461 galaxies.

After apparent magnitude, the two most important cuts in our sample
are on surface-brightness and absolute magnitude. It has been
demonstrated that unless the shapes of the luminosity and
surface-brightness functions change dramatically at low
surface-brightnesses, the surface-brightness cuts exclude at most
5~percent of the luminosity density (Blanton et al 2001).  The
Schechter function fit to the SDSS luminosity function (from Blanton
et al 2001 but corrected to the $\fixr$-band) has 90~percent of the
luminosity density in the range $-23.5<\fixMr<-17.5~\mathrm{mag}$.  At
least 85~percent of the luminosity density of the Universe is
therefore in our sample; it can be used to make reliable estimates of
the fraction of the cosmic luminosity density found in galaxies of
different types.

\subsection{Density contributions}

To compute number and luminosity densities, it is necessary to compute
the number-density contribution $1/\Vmax$ for each galaxy, where
$\Vmax$ is the volume covered by the survey in which this galaxy could
have been observed, accounting for the flux, surface brightness, and
redshift limits as a function of angle (Schmidt 1968). This volume is
calculated as follows:
\begin{equation}
\Vmax = \frac{1}{4\pi}\,\int\mathrm{d}\Omega\,f(\theta,\phi)\, 
\int_{z_{\mmin(\theta,\phi)}}^{z_{\mmin(\theta,\phi)}}
\mathrm{d}z\,\frac{dV}{dz} \;,
\end{equation}
where $f(\theta,\phi)$ is described below and
$z_{\mmin}(\theta,\phi)$ and $z_{\mmax}(\theta,\phi)$
are defined by:
\begin{eqnarray}\displaystyle
z_{\mmin}(\theta,\phi) &=& \mmax(
z_{m,\mmin}(\theta,\phi), z_{\mu,\mmin}(\theta,\phi),
0.02)\cr
z_{\mmax}(\theta,\phi) &=& \mmin(
z_{m,\mmax}(\theta,\phi), z_{\mu,\mmax}(\theta,\phi),
0.22) \;.
\end{eqnarray}
The flux limits $m_{r,\mmin}(\theta,\phi)$ (equal to 14.5~mag across
the survey) and $m_{r,\mmax}(\theta,\phi)$ (approximately 17.77~mag,
but slightly variable as noted above) implicitly set
$z_{m,\minmax}(\theta,\phi)$ by:
\begin{eqnarray}\displaystyle
m_{r,\minmax}(\theta,\phi) &=& \fixMr +
\mathrm{DM}(z_{m,\minmax}(\theta,\phi))\cr
& & + K_\fixr(z_{m,\minmax}(\theta,\phi)) \;.
\end{eqnarray}
The surface brightness limits $\mu_{r,\mmin}(\theta, \phi)$ (in
practice, too high surface-brightness to matter for this sample) and
$\mu_{r,\mmax}(\theta,\phi)$ (approximately 24.5~mag but variable
in a known way across the survey) implicitly set
$z_{\mu,\minmax}(\theta,\phi)$ by:
\begin{eqnarray}\displaystyle
\mu_{r,\minmax}(\theta,\phi) &=& \mu_{r} +
10 \log_{10} (1+z_{\mu,\minmax}(\theta,\phi))\cr
& & + K_\fixr(z_{\mu,\minmax}(\theta,\phi)) \;.
\end{eqnarray}
In practice, the surface brightness limits only rarely affect the
$\Vmax$ determination.

The function $f(\theta,\phi)$ is the SDSS sampling fraction of
galaxies as a function of position on the sky. The total sampling rate
of galaxies is computed separately for each region covered by a unique
set of survey tiles.  The nomenclature of the 2dF
(Percival et al 2001) is adopted; each such region is a ``sector''
(which corresponds identically an ``overlap region'' in Blanton et al
2001). That is, in the case of two overlapping tiles, the sampling is
calculated separately in three sectors: the sector covered only by the
first tile, the sector covered only by the second tile, and the sector
covered by both.  Each position on the sky is thereby assigned a
sampling rate $f(\theta,\phi)$ equal to the number of redshifts of
galaxy targets obtained in the region divided by the number of galaxy
targets in the region. In regions covered by a single tile, typically
$0.85<f<0.9$; in multiple plate regions, typically $f>0.95$. These
completenesses average to 0.92.

\section{Distribution in ``photometric space''}

In this Section, the distribution of galaxies in a measurement space
made up of colors, surface brightness, and concentration of the radial
profile is displayed and discussed.  For the purposes of this section,
the surface brightness is defined to be ($-2.5\,\log_{10}$ of) half of
the Petrosian flux in the $\fixr$ band, divided by
$\pi\,\theta_{50}^2$, and the concentration $c$ is defined to be the
ratio $\theta_{90}/\theta_{50}$.

\subsection{Number densities}

\begin{figure}
\plotone{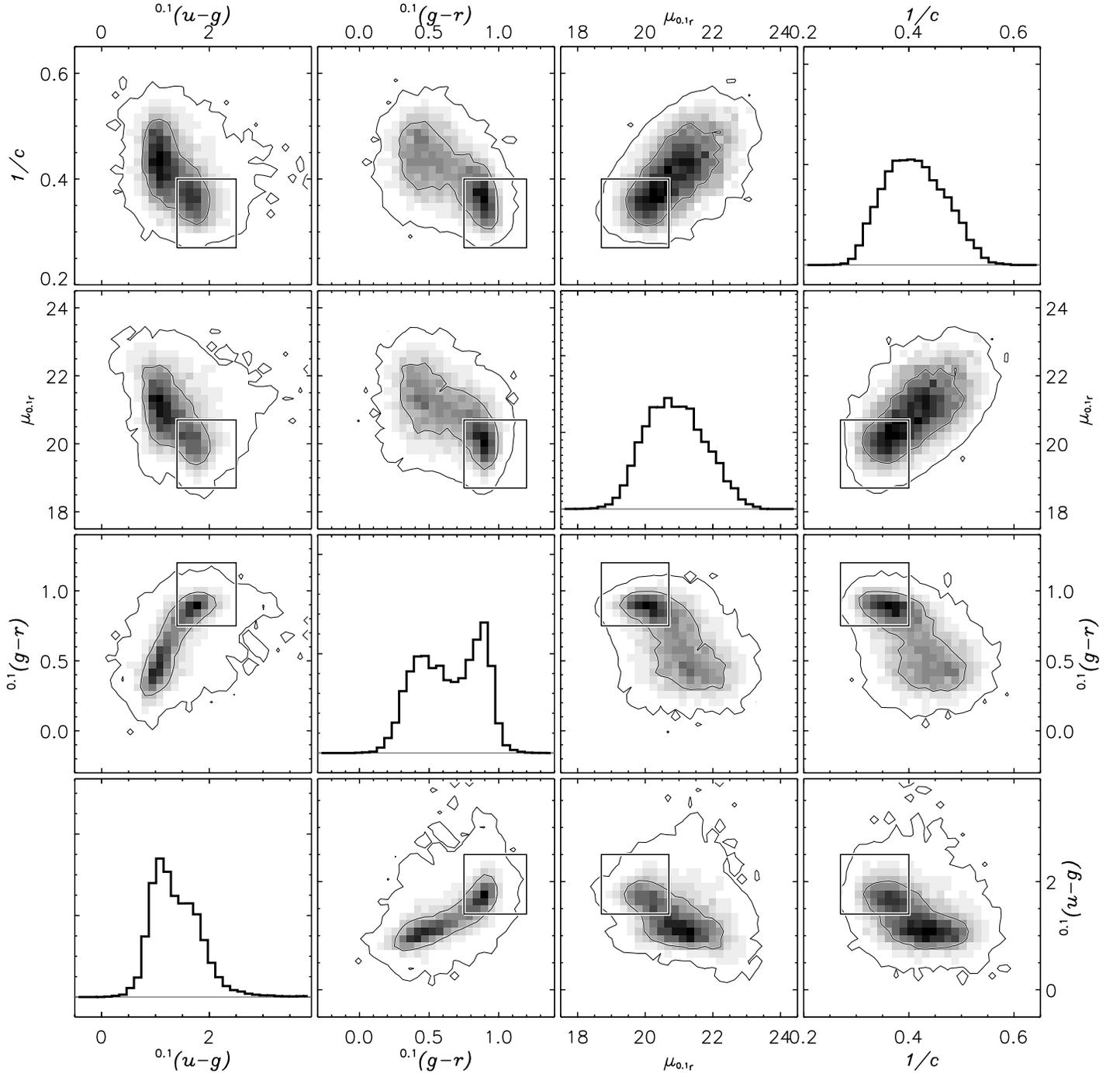}\figcaption[number]{The distribution of galaxy
number density in the four-dimensional photometric space of
fixed-frame color $\fixug$, fixed-frame color $\fixgr$, central
surface brightness $\fixmur$, and inverse concentration $1/c$.  All
four photometric properties and their measurement are described in the
text.  Recall that the fixed-frame colors are fixed at redshift 0.1
not 0.  The four panels on the diagonal show the one-dimensional
distributions of each of the four photometric properties.  Each
histogram bin shows the total number density of galaxies in that bin,
computed with the selection volumes $\Vmax$.  The twelve off-diagonal
panels show the two-dimensional distributions for each pair of
parameters.  The shading of each pixel is proportional to the number
density of galaxies with properties in that pixel, computed with the
selection volumes $\Vmax$.  Projections of the four-dimensional
rectangular box that encloses the red galaxy population are
superimposed on the off-diagonal panels.
\label{fig:number}}
\end{figure}

For any particular subset of galaxies chosen to have specific physical
properties, the sum of the inverse selection volumes $1/\Vmax$ for the
members of the subset is a measure of the number density of galaxies
with those physical properties.  Figure~\ref{fig:number} shows
projections of the number density as a function of position in the
four-dimensional photometric space spanned by fixed-frame color
$\fixug$, fixed-frame color $\fixgr$, fixed-frame surface brightness
$\fixmur$ and concentration $1/c$.

There are two separated density maxima apparent in
Figure~\ref{fig:number}, one corresponding to red, concentrated, high
surface-brightness galaxies, and one corresponding to bluer, less
concentrated, lower surface-brightness galaxies.  For clarity, a
four-dimensional box containing the redder peak has been superimposed
on Figure~\ref{fig:number}.  The box was chosen by hand to be
restrictive but still encompass most of the red peak.  The box is
defined by the limits:
\begin{eqnarray}\displaystyle
1.40 < & \fixug  & < 2.50 \cr
0.75 < & \fixgr  & < 1.20 \cr
18.7 < & \fixmur & < 20.7 \cr
0.27 < & (1/c)   & < 0.40
\end{eqnarray}
where the colors are measured in mag and the central surface
brightness $\fixmur$ is measured in units of mag in one arcsec$^2$.
We refer to the galaxies inside this four-dimensional box, ie,
galaxies that simultaneously satisfy all four photometric cuts, as
``red-type'' galaxies.  The separation of red-type and non-red-type
galaxies is very similar to that shown previously to be highly
correlated with galaxy morphological type (Strateva et al 2001).

The integrated number density (ie, total of $1/\Vmax$ values) of
red-type galaxies (ie, galaxies in the four-dimensional box) in the
sample is 20~percent of the total number density in \lsssample.

\subsection{Luminosity densities}

\begin{figure}
\plotone{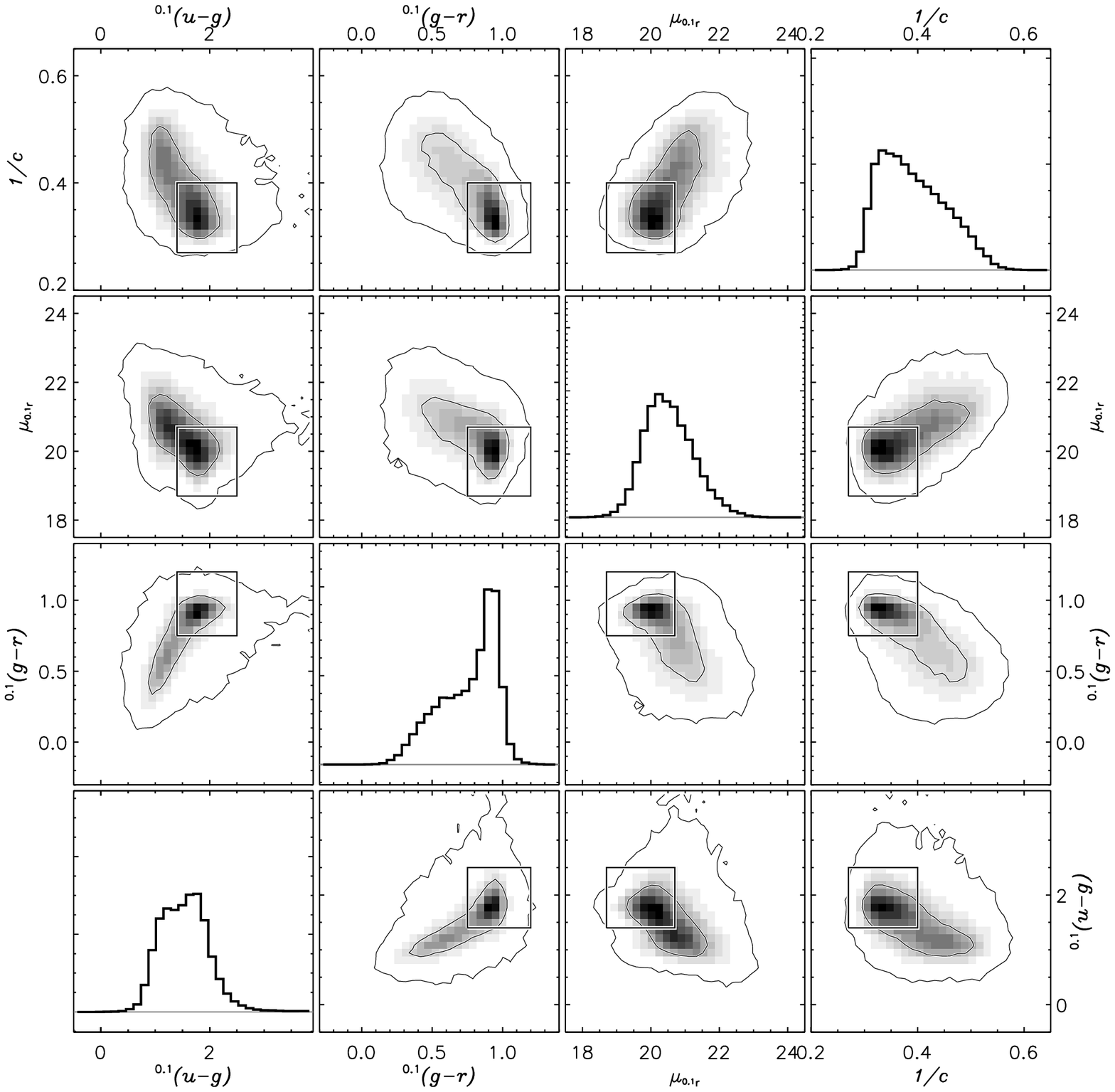}\figcaption[luminosity]{The distribution of
$\fixi$-band galaxy luminosity density in the four-dimensional
photometric space.  Axes and box as in Figure~\ref{fig:number}.
Recall that the fixed-frame colors are fixed at redshift 0.1, not 0.
\label{fig:luminosity}}
\end{figure}

Every galaxy in the sample has a density contribution $1/\Vmax$ and an
$\fixi$-band luminosity $\fixLi$, computed with $\fixi_\mathrm{pet}$
and the luminosity distance (eg, Hogg 1999), so every galaxy has a
luminosity density contribution $\fixLi/\Vmax$.
Figure~\ref{fig:luminosity} shows the total luminosity density of the
sample as a function of position in the photometric space.

The integrated luminosity density of red-type galaxies (ie, galaxies
in the four-dimensional box) in the sample is 38~percent of the total
luminosity density.

\section{Discussion}

We have identified a class of ``red-type'' galaxies in the Sloan
Digital Sky Survey on the basis of their colors, surface brightnesses,
and radial profiles.  We have shown that roughly two fifths of the
total cosmic galaxy luminosity density in the visual is produced by
these red galaxies.

The high central surface brigtnesses and strong radial concentrations
of the red galaxies suggest that they are red by dint of age rather
than dust extinction.  Galaxies outside the red-type box will not
enter the box by becoming dustier, because the dust that makes them
redder will also further reduce their surface brightnesses.

It is conventional (though unnecessary in the present work) to type
galaxies morphologically.  A sample of 100 galaxies at $z<0.1$ was
chosen from the red-type sample for morphological classification.  The
galaxies were chosen randomly, with probability proportional to number
density contribution $1/\Vmax$ (re-computed in the face of the
additional $z<0.1$ cut).  Three authors (MB, DWH, IS) independently
classified these 100 galaxies into the coarse morphological sequence
E, S0, Sa, Sb/Sc, Sd/Irr.  Classification was based on
$\nu\,f_\nu$-weighted averages of the $r$ and $i$-band images, viewed
in black and white with variable stretch on computer screens.  For
95~percent of the sample, each pair of classifiers agree to within one
morphological class in the crude sample.  Across all classifiers, 65
to 77~percent of the number-density weighted sample is in the E and S0
classes.  At most 5 percent of the sample is later than Sa in
morphological type.  Note that none of our conclusions depend on this
by-eye morphological classification.


That red galaxies contribute a large fraction of the luminosity
density is remarkable because bluer stellar populations produce much
more light per unit mass than redder.  Typical red-type galaxies in
the sample have $\fixug=1.8~\mathrm{mag}$ and
$\fixgr=0.9~\mathrm{mag}$; while typical galaxies outside the red-type
region have $\fixug=1.0~\mathrm{mag}$ and $\fixgr=0.5~\mathrm{mag}$
(recall that these are not zero-redshift colors).  Conventional
stellar population models suggest that these two colors differ in
stellar mass-to-light ratio by a factor of two or more (Bruzual \&
Charlot 1993; Kauffmann et al 2002), with the specific value depending
on the assumed star formation history in each case.  (Patchy or
inhomogeneous dust extinction can modify the specific mass-to-light
ratios, but dust obeys the same general trend, ie, that redder
populations have higher mass-to-light ratios than blue.)  The natural
interpretation is that more than 50~percent of the stars in the
Universe are in red galaxies.

This result is conservative, because we have only considered a very
restricted set of red-type galaxies: those that simultaneously satisfy
all four of our cuts on color, surface brightness and concentration.
There are certainly significant red stellar populations contained in
the centers of otherwise blue spiral galaxies, and old stellar
populations contribute significantly to the stellar masses of almost
all galaxies in the $z<0.25$ Universe.  A very large fraction of all
stars are in old stellar populations.

In conventional models of stellar population evolution,
$\fixug=1.8~\mathrm{mag}$ and roughly Solar metallicity corresponds to
an age greater than about $7~\mathrm{Gyr}$ (eg, Bruzual \& Charlot
1993), or a formation redshift of $z_\mathrm{form}>1$ (if the Universe
is $14~\mathrm{Gyr}$ old and the median redshift of the sample is
$z=0.1$).  Such derived ages depend strongly on metallicity, and red
galaxies are known to have abundance ratios different from Solar (they
are alpha-enhanced; eg, Worthey et al 1992), so it is dangerous to
take absolute measures of stellar population ages seriously.  However,
taken at face value, it suggests that the cosmic star formation rate
was extremely high at early times.  Although much is known about star
formation history since $z\sim 1$ (see references in Hogg 2001),
little is known about $z>1$.  In a minimal model, the comoving
volume-averaged star formation rate at $z< 1$ can be taken to go as
$(1+z)^{3.1}$ (Hogg 2001) and at $z> 1$ can be imagined to be roughly
constant and equal to the rate at $z\sim 1$, then roughly $65$~percent
of stars would be expected to be older than $8~\mathrm{Gyr}$ today.
This minimal model is consistent with our results.  Given the large
numbers of old stars residing in red galaxies in the SDSS,
supplemented by the old stars residing in bluer galaxies, it is hard
see how $z> 1$ comoving star formation rate densities could have been
significantly below $z\sim 1$ values for any significant fraction of
cosmic time.

\acknowledgements It is a pleasure to thank Daniel Eisenstein, Douglas
Finkbeiner, Jim Gunn, Jim Peebles, Sam Roweis, David Spergel, Michael
Strauss, Christy Tremonti and David Weinberg for stimulating
discussions, useful software, and help with the literature.  This
research made use of the NASA Astrophysics Data System.  DWH and MB
are partially supported by NSF (grant 0101738) and NASA (grant
NAG5-11669); IS, NAB, ZI, RL, GRK and DJS thank Princeton University
for generous support.

Funding for the creation and distribution of the SDSS Archive has been
provided by the Alfred P. Sloan Foundation, the Participating
Institutions, the National Aeronautics and Space Administration, the
National Science Foundation, the U.S. Department of Energy, the
Japanese Monbukagakusho, and the Max Planck Society. The SDSS Web site
is http://www.sdss.org/.

The Participating Institutions are The University of Chicago,
Fermilab, the Institute for Advanced Study, the Japan Participation
Group, The Johns Hopkins University, Los Alamos National Laboratory,
the Max-Planck-Institute for Astronomy, the Max-Planck-Institute for
Astrophysics, New Mexico State University, Princeton University, the
United States Naval Observatory, and the University of Washington.

\clearpage

\clearpage


\begin{thebibliography}{}
\bibitem{blanton01}
Blanton,~M.~R. et~al 2001, \aj, 121, 2358
\bibitem{blanton02tiling}
Blanton,~M.~R. et~al 2002a, \aj, submitted
\bibitem{blanton02kcorrect}
Blanton,~M.~R. et~al 2002b, in preparation
\bibitem{bruzual93}
Bruzual~A.,~G. \& Charlot,~S. 1993, \apj, 405, 538
\bibitem{connolly97}
Connolly,~A.~J., Szalay,~A.~S., Dickinson,~M., Subbarao,~M.~U.,
\& Brunner,~R.~J. 1997, \apj, 486, L11
\bibitem{kormendy89}
Kormendy,~J. \& Djorgovski,~S. 1989, \araa, 27, 235
\bibitem{fukugita98}
Fukugita,~M., Hogan,~C.~J. \& Peebles,~P.~J.~E. 1998, \apj, 503, 518
\bibitem{fukugita96}
Fukugita,~M., Ichikawa,~T., Gunn,~J.~E., Doi,~M., Shimasaku,~K., \&
Schneider,~D.~P. 1996, \aj, 111, 1748
\bibitem{gunn98}
Gunn,~J.~E. et~al 1998, \aj, 116, 3040
\bibitem{hogg99cosm}
Hogg,~D.~W. 1999, astro-ph/9905116
\bibitem{hogg01}
Hogg,~D.~W. 2001, astro-ph/0105280
\bibitem{kauffmann02}
Kauffmann,~G. et~al 2002, \mnras, submitted (astro-ph/0204055)
\bibitem{lilly96}
Lilly,~S.~J., Le Fevre,~O., Hammer,~F., \& Crampton,~D. 1996, \apjl, 460, L1
\bibitem{lupton01}
Lupton,~R.~H., Gunn,~J.~E., Ivezi\'{c},~Z., Knapp,~G.~R., Kent,~S., \&
Yasuda,~N. 2001, ASP Conf.~Ser.~238: Astronomical Data Analysis
Software and Systems X, 10, 269
\bibitem{oke83}
Oke,~J.~B. \& Gunn,~J.~E. 1983, \apj, 266, 713
\bibitem{pei95}
Pei,~Y.~C. \& Fall,~S.~M. 1995, \apj, 454, 69
\bibitem{percival01a}
Percival,~W.~J. et~al 2001, \mnras, 327, 1297
\bibitem{petrosian76}
Petrosian,~V. 1976, \apj, 209, L1
\bibitem{pier02}
Pier,~J.~R., 2002, \aj, submitted
\bibitem{roberts94}
Roberts,~M.~S. \& Haynes,~M.~P.\ 1994, \araa, 32, 115
\bibitem{schlegel98}
Schlegel,~D.~J., Finkbeiner,~D.~P. \& Davis,~M. 1998, \apj, 500, 525
\bibitem{schmidt68}
Schmidt,~M., \apj, 151, 393
\bibitem{smith02}
Smith,~J.~A., Tucker,~D.~L. et~al 2002, \aj, in press
\bibitem{strateva01}
Strateva,~I. et~al 2001, \aj, 122, 1861
\bibitem{strauss02}
Strauss,~M.~A. et~al 2002, \aj, submitted
\bibitem{worthey92}
Worthey,~G., Faber,~S.~M., \& Gonzalez,~J.~J., 1992, \apj, 398, 69
\bibitem{york00}
York,~D.~G. et~al 2000, \aj, 120, 1579
\end{thebibliography}
\end{document}